\documentclass[aps,groupedaddress]{revtex4}
\usepackage{graphicx}
\usepackage{amsmath}

\newcommand{\half}{\frac{1}{2}}
\DeclareMathOperator{\Li}{Li}

\begin{document}
\bibliographystyle{apsrev}


\title{Phase separation of penetrable core mixtures}


\author{R. Finken}%
\email{rf227@cam.ac.uk}
\author{J.-P. Hansen}
\author{A. A. Louis}

\affiliation{Department of Chemistry\\
University of Cambridge\\
Cambridge CB2 1EW (UK)}


\date{\today}

\begin{abstract}
A  two-component system of penetrable particles interacting via a
gaussian core potential is considered, which may serve as a crude
model for binary polymer solutions. The pair structure and
thermodynamic properties are calculated within the random phase
approximation (RPA) and the hypernetted chain (HNC) integral
equation. The analytical RPA predictions are in semi-quantitative
agreement with the numerical solutions of the HNC approximation, which
itself is very accurate for gaussian core systems. A fluid-fluid phase
separation is predicted to occur for a broad range of potential
parameters. The pair structure exhibits a nontrivial clustering
behaviour of the minority component. Similiar conclusions hold for the
related model of parabolic core mixtures, which is frequently used in
dissipative particle dynamics (DPD) simulations.
\end{abstract}

\maketitle


\section{Introduction}
\label{sec:introduction}

Demixing of binary or multicomponent mixtures is a very common
phenomenon observed in a broad range of molecular fluids
\cite{ref1}, polymer solutions and blends
\cite{ref2a,ref2b}, or colloidal dispersions \cite{ref3,ref4}.
Phase separation is generally associated with differences in the attractive
interactions between particles of different chemical species. In
polymer solutions these differences are usually embodied in the Flory
$\chi$-parameter \cite{ref2a}, which controls the competition between
the entropy of mixing and the total interaction energy, at least within
a mean-field picture. On the other hand, in multi-component colloidal systems like binary
dispersions involving colloidal particles of very different sizes, or
mixtures of colloidal particles and non-adsorbing polymer, phase
separation can be driven by purely repulsive, excluded volume
interactions. By mapping the initial multi-component system onto an
{\em effective}\/ one-component system involving
only the bigger colloidal particles, the largely entropy-driven
demixing can be understood in terms of {\em attractive}\/ depletion
interactions induced between the large particles by the smaller
species (the ``depletant'') \cite{ref4}. It should however be kept in
mind that the initial bare interactions are purely repulsive, albeit
strongly non-additive, as in the highly simplified Asakura-Oosawa model
for colloid-polymer mixtures \cite{ref5}. In the case of fully
additive hard sphere mixtures, phase separation has been predicted for
sufficiently large size ratios \cite{ref6}, but it is now generally
believed that the fluid-fluid demixing is metastable, and preempted by
freezing \cite{ref7}. A significant degree of positive non-additivity
of the core radii $R_{\mu \nu}$ (whereby $R_{12} =
(R_{11}+R_{22})(1+\Delta)/2$, with $\Delta >0$) is required to observe
a stable demixing transition in the fluid phase \cite{ref8}.

Effective interactions between the centres of mass of fractal objects,
like linear polymer coils \cite{ref9,ref10,ref11,ref12}, star polymers
\cite{ref13} or dendrimers \cite{ref14}, obtained by averaging over
individual monomer degrees of freedom, are now known to be very
``soft''. More specifically the effective pair potential diverges only
logarithmically for overlapping star polymers \cite{ref13}, while
remaining finite, of the order of $1-2 k_B T$, for linear polymers in
good solvent \cite{ref9,ref10,ref11,ref12}. This observation
has stimulated the investigation of simple models, like finite
repulsive step potentials \cite{ref15}, or the gaussian core potential
\cite{ref16,ref17,ref18}, which was first introduced by Stillinger, in
a somewhat different context \cite{ref19}, namely
\begin{equation}
  \label{eq:1}
  v(r) = \epsilon \exp\left(-r^2/R^2\right),
\end{equation}
where $\epsilon$ is the energy scale, while $R$ determines the range
of the effective potential. It is worth stressing that the ``gaussian
core'' model is unrelated to the ``gaussian molecule'' model, which
was extensively studied by Michael Fisher and collaborators
\cite{ref26}. In the latter model it is the Mayer $f$-function, rather
than the pair potential, which has a gaussian shape. 

Simple, penetrable particle models are also widely used in highly
coarse-grained simulations of large-scale phenomena within the
so-called ``dissipative particle dynamics'' (DPD) method
\cite{ref20,ref21}. In DPD, effective interactions between penetrable
fluid ``particles'' are frequently modelled by a simple parabolic
potential \cite{ref21,ref22}:
\begin{equation}
  \label{eq:2}
  v(r) = \begin{cases}
    \epsilon (1-r/R)^2;& r < R\\
    0; & r \ge R.
  \end{cases}
\end{equation}
It has recently been realized that binary mixtures of particles with
penetrable cores, which interact via generalizations of the gaussian
and parabolic potentials \eqref{eq:1} and \eqref{eq:2}, involving
different energy scales $\epsilon$ and radii $R$ for the various
species, may phase-separate over appropriate ranges of these
parameters. Spinodal instability was first shown to occur for the
gaussian core model \eqref{eq:1} within the random phase approximation
(RPA) \cite{ref17} and binodals as well as interfacial properties were
then calculated within the same approximation \cite{ref18}. Similarly
Gibbs ensemble Monte Carlo simulations have very recently shown that
binary systems of soft particles interacting via the parabolic
potential \eqref{eq:2} phase separate beyond a critical degree of
enhanced repulsion between particles of different species, in agreement with
Flory-like mean field considerations \cite{ref22}.

In this paper we systematically extend our earlier results for the
gaussian core model \cite{ref17} and investigate the range of validity
of the RPA by detailed calculations of the pair structure,
thermodynamics, and the resulting phase coexistence curve within the
much more accurate hypernetted chain (HNC) approximation. The
break-down of the RPA is quantified in the physically relevant regime
where $\rho R^3 \simeq 1$ and $\epsilon \simeq k_B T$, which would
correspond to the cross-over from dilute to semi-dilute regimes of the
underlying binary polymer solution. The RPA continues to provide
reliable first estimates at higher densities.

\section{RPA and HNC}
\label{sec:rpa-hnc}

The model under consideration is the binary gaussian core model (GCM)
already introduced in references \cite{ref17} and \cite{ref18}. It
consists of $N_1$ particles of ``radius'' $R_1$ and $N_2$ particles of
``radius'' $R_2$ in a volume $V$. The total number density is $\rho =
(N_1 + N_2) / V$, while the concentrations of the two species are $x =
N_2 / N$ and $1-x = N_1 / N$, respectively. The
pair potentials are
\begin{equation}
  \label{eq:3}
  v_{\mu \nu}(r) = \epsilon_{\mu \nu} e^{-\left(r/R_{\mu \nu}\right)^2},
\end{equation}
which introduce three length and three energy parameters: $R_{11}$,
$R_{22}$, $R_{12}, \epsilon_{11}, \epsilon_{12}$ and $\epsilon_{22}$.
If $R_{11}$ is chosen as unit of length, the system is entirely
specified by the 5 dimensionless parameters $R_{12} / R_{11}$, $R_{22}
/ R_{11}$, $\epsilon^*_{11} = \beta \epsilon_{11}$, $\epsilon^*_{12} =
\beta \epsilon_{12}$, $\epsilon^*_{22} = \beta \epsilon_{22}$, where
$\beta = 1/ (k_B T)$. For a fixed set of dimensionless parameters the
reduced Helmholtz free energy per particle, $f = F / (N k_B T)$, is a
function only of the intensive variables $\rho$ and $x$; this may be
split into the ideal gas , ideal mixing and excess (non-ideal)
contributions:
\begin{subequations}
  \label{eq:4}
  \begin{align}
    f(\rho,x) &= f_{\text{id}}(\rho) + f_{\text{mix}}(x) +
    f_{\text{ex}}(\rho,x)\\
    &= \ln(\rho \Lambda^3) - 1 + x \ln x + (1-x) \ln (1-x) +
    f_{\text{ex}}(x,\rho),
  \end{align}
\end{subequations}
where $\Lambda$ is an irrelevant de Broglie thermal wavelength. The
equation of state $\beta P / \rho$, and the chemical potentials
$\mu_\nu,\, (\nu = 1,2)$ follow from the standard thermodynamic
relations:
\begin{subequations}
  \label{eq:5}
  \begin{align}
    \frac{\beta P}{\rho} &= \rho \left(\frac{\partial f}{\partial \rho}
    \right)_x\\
    \beta \mu_1 &= \left(\frac{\partial \rho f}{\partial
        \rho}\right)_x - x \left(\frac{\partial f}{\partial x}\right)_\rho\\
    \beta \mu_2 &= \left(\frac{\partial \rho f}{\partial \rho}\right)_x
    + (1-x) \left(\frac{\partial f}{\partial x}\right)_\rho
  \end{align}
\end{subequations}
and may likewise be split into ideal and excess contributions. Within
RPA and HNC, these excess contributions may be easily expressed in
terms of the usual total and direct correlation functions $h_{\mu
  \nu}(r)$ and $c_{\mu \nu}(r)$, which are related by the familiar
Ornstein-Zernike (OZ) relations \cite{ref23}; these may be used to
express the Fourier transforms (FT) $\hat{h}_{\mu \nu}(k)$ of the
total correlation functions in terms of the FT $\hat{c}_{\mu \nu}(k)$
of the direct correlation functions:
\begin{subequations}
  \label{eq:6}
  \begin{align}
    \hat{h}_{11}(k) &= \frac{1}{\Delta(k)}
    \left[\hat{c}_{11}(k)(1-\rho_2 \hat{c}_{22}(k)) + \rho_2
      \hat{c}_{12}^2(k)\right],\\
    \hat{h}_{12}(k) &= \frac{1}{\Delta(k)} \hat{c}_{12}(k),\\
    \hat{h}_{22}(k) &= \frac{1}{\Delta(k)}
    \left[\hat{c}_{22}(k)(1-\rho_1 \hat{c}_{11}(k)) + \rho_1
      \hat{c}_{12}^2(k)\right],
  \end{align}
\end{subequations}
where $\rho_1 = \rho (1-x), \rho_2 = \rho x,$ and 
\begin{equation}
  \label{eq:7}
  \Delta(k) = [1-\rho_1 \hat{c}_{11}(k)][1-\rho_2 \hat{c}_{22}(k)] - \rho_1 \rho_2
  \hat{c}_{12}^2(k).
\end{equation}
These OZ relations must be supplemented by closure relations. 

The RPA amounts to simply identifying the $c_{\mu \nu}(r)$ with their
asymptotic (large $r$) behaviour:
\begin{align}
  \label{eq:8}
  c_{\mu \nu}(r) &= - \beta v_{\mu \nu}(r)\nonumber\\
  &= - \epsilon_{\mu \nu}^* e^{-\left(r/R_{\mu \nu}\right)^2}
\end{align}
Substitution of the FT's of \eqref{eq:8} into the relations
\eqref{eq:6} and \eqref{eq:7} yields explicit expressions for the
$\hat{h}_{\mu \nu}(k)$, which may be transformed back to obtain the
$h_{\mu \nu}(r)$. Detailed analytical expressions are given in
Appendix \ref{sec:solut-viri-route} in the special case of mixtures
with equal core radii $R_{11} = R_{22} = R_{12}= R$.

The HNC closure relations are \cite{ref23}:
\begin{subequations}
  \label{eq:9}
  \begin{align}
    g_{\mu \nu}(r) &= 1 + h_{\mu \nu}(r)\\
    &= \exp{\{-\beta v_{\mu \nu}(r) + \gamma_{\mu \nu}(r)\}}\nonumber\\
    \gamma_{\mu \nu}(r) &= h_{\mu \nu}(r) - c_{\mu \nu}(r)
  \end{align}
\end{subequations}
An iterative procedure must be used to solve the coupled equations
\eqref{eq:6} and \eqref{eq:9} numerically. Convergence is easily
achieved for the binary GCM model over most of the potential parameter
space. In fact the HNC approximation becomes exact in the high density
limit ($\rho R^3 \rightarrow \infty$), and is extremely
accurate for densities $\rho R^3 \approx 1$, as shown earlier in the
one-component case \cite{ref16,ref17}. Pair structure data will be
examined in greater detail in section 5.

Knowledge of the pair correlation functions allows direct access to
thermodynamics, via the compressibility or virial routes \cite{ref23}.
Being approximate, the RPA and HNC closures are not thermodynamically
consistent, i.\ e.\ the two routes lead to different answers. In the
case of the GCM the two theories become strictly thermodynamically
consistent only in the high density limit, \cite{ref16,ref17}.

Within the RPA, the compressibility route is equivalent to a simple
mean-field ansatz for the free energy \cite{ref16,ref17}. This leads
immediately to the following analytic expression for the excess part
of the reduced free energy \cite{ref17}:
\begin{subequations}
  \label{eq:10}
  \begin{align}
    f_{\text{ex}}^{\text{C}}(x,\rho) &= \half \rho V_0(x)\\
    &= \half \rho \sum_{\nu} \sum_\mu x_\mu x_\nu V_{\mu \nu}\\
    V_{\mu \nu} &= \beta \hat{v}_{\mu \nu}(k = 0) = \int \beta v_{\mu
      \nu}(r) d\mathbf{r},
  \end{align}
\end{subequations}
where $x_1 = (1-x), x_2 = x$. In the special case of the GCM:
\begin{equation}
  \label{eq:11}
  V_{\mu \nu} = \pi^{3/2} \epsilon^*_{\mu \nu} R_{\mu \nu}^3
\end{equation}

The pressure and chemical potentials $\mu_\nu$ follow directly from
eqns. \eqref{eq:5} (the superscript C refers to the compressibility
route):
\begin{subequations}
  \label{eq:12}
  \begin{align}
    \frac{\beta P^{\text{C}}}{\rho} &= 1 + \frac{\rho}{2} \sum_{\mu}
    \sum_{\nu} x_{\mu} x_\nu V_{\mu \nu}\label{eq:12a}\\
    \label{eq:12b}
    \beta \mu^{\text{C}}_1 &= \ln(\rho \Lambda^3 (1-x)) + V_{11} \rho
    (1-x) + V_{12} \rho x\\
    \label{eq:12c}
    \beta \mu^{\text{C}}_2 &= \ln(\rho \Lambda^3 x) + V_{12} \rho (1-x)
    + V_{22} \rho x
  \end{align}
\end{subequations}
The virial route is considerably more arduous. The pressure follows
from the virial theorem:
\begin{align}
  \label{eq:13}
  \frac{\beta P^{\text{V}}}{\rho} &= 1 + \frac{\rho}{2} \sum_\mu
  \sum_\nu x_\mu x_\nu V_{\mu \nu}\\
  &\quad - \frac{2 \pi}{3} \rho \sum_\mu \sum_\nu x_\mu x_\nu \int_0^\infty r^3
  \frac{d \beta v_{\mu \nu}(r)}{dr} h_{\mu \nu}(r) dr\nonumber
\end{align}
In the case of the symmetric mixture ($R_{11} = R_{22} = R_{12} = R$;
$V_{11} = V_{22}$) the RPA pair correlation functions obtained in
Appendix A may be used to yield the following expression (the
superscript V referring to the virial route):
\begin{align}
  \label{eq:14}    \frac{\beta P^{\text{V}}}{\rho} &= \frac{\beta P^{\text{C}}}{\rho} -
  \frac{z_2 x (1-x) \Delta}{\pi^{3/2} R^3}
  \aleph\left(-\frac{\rho}{z_1}\right)\nonumber \\
  &\qquad - \frac{z_1 x (1-x) \Delta}{\pi^{3/2} R^3}
  \aleph\left(-\frac{\rho}{z_2}\right),
\end{align}
where the functions $\aleph(s)$ and the roots $z_1$ and
$z_2$ are defined in appendix \ref{sec:solut-viri-route} (eqns.~\eqref{eq:b4a} and
\eqref{eq:b4b}). The resulting reduced free energy per particle is obtained by
integrating the equation of state \eqref{eq:14} with respect to
density:
\begin{equation}
  \label{eq:15}
  f^{\text{V}}(x, \rho) = f^{\text{C}}(x, \rho) + \frac{1}{2
    \pi^{3/2} \rho R^3} \left[\Li_{5/2}(\rho/z_1)
    + \Li_{5/2}(\rho/z_2) + \rho V_{11}\right],
\end{equation}
where the polylogarithm $\Li_{5/2}$ is defined in equation
\eqref{eq:a12}. Unfortunately this expression becomes singular for
densities larger than the spinodal densities obtained by the
compressibility route. This mathematical artifact discussed in
Appendix \ref{sec:solut-viri-route} does not allow us to use
expressions \eqref{eq:14} and \eqref{eq:15} to construct a phase
diagram.

\begin{figure}[htbp]
  \begin{center}
    \includegraphics[angle=-90,width=14cm]{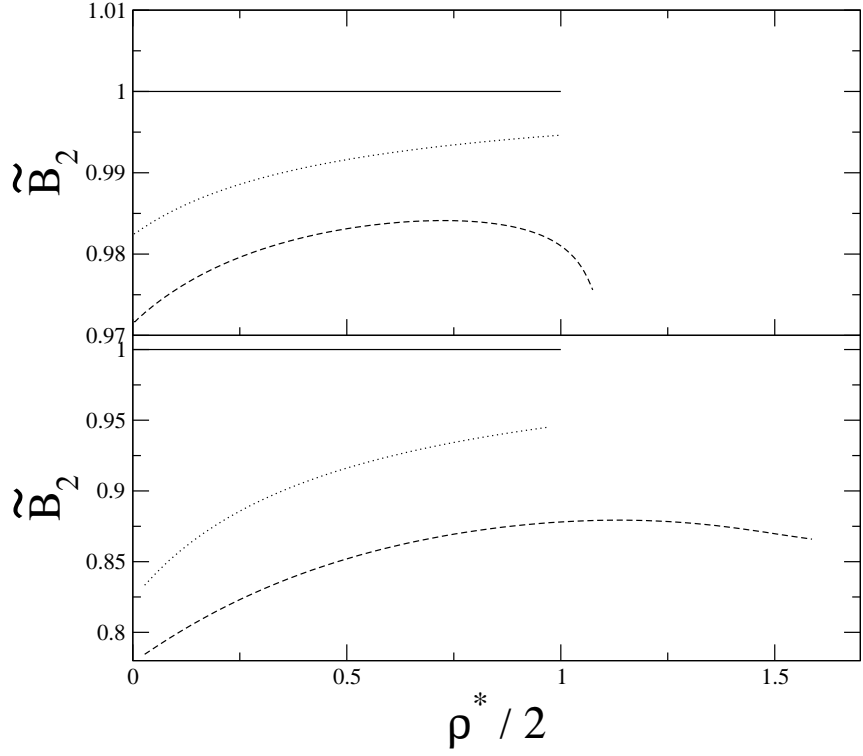}
    \caption{The normalised effective second virial coefficient
      $\tilde{B}_2 = 4 (\beta P - \rho)/ (3 \rho^2 V_{11})$ is plotted versus the
      normalized density $\rho^* / 2$ at fixed concentration $x =
      0.5$ for RPA-C, RPA-V and HNC . In the upper plot, we have
      $\epsilon^*_{11} = \epsilon^*_{22} = 0.1, \epsilon^*_{12} = 0.2$,
      while in the lower plot $\epsilon^*_{11} = \epsilon^*_{22} = 1.0,
      \epsilon^*_{12} = 2.0$. Both systems have the reduced critical
      density $\rho^*_c = 2$ in RPA-C.
      The RPA-C route predicts
      $\tilde{B}_2 = 1$ at all densities (solid line). The RPA-V
      values are represented by the dotted lines, the HNC data by the
      dashed lines up to the density beyond which convergence fails.
      The agreement between the three routes to the equation of state
      is seen to improve with increasing density, except in the
      vincinity of the critical density where the results diverge again.}
    \label{fig:1}
  \end{center}
\end{figure}

The lack of thermodynamic consistency of the RPA is illustrated
in Figure~\ref{fig:1}, which compares the compressibility and virial
equations of state as a function of density, for a symmetric mixture;
the virial pressures are always lower than their compressibility
counterparts, and closer to the nearly exact HNC results, to which we
now turn.

The HNC approximation is very nearly thermodynamically consistent in
practice, at least for temperatures and densities relevant for polymer
solutions \cite{ref17}. Hence we have calculated thermodynamic
properties within the more convenient virial route with the pressure
given by the standard relation \eqref{eq:13}, while the chemical
potentials may also be directly expressed in terms of the direct and
total correlation functions, according to \cite{ref24}:
\begin{align}
  \label{eq:16}
  \beta \mu_\nu &= \ln(\rho_\nu \Lambda^3) + \sum_\mu \left\{
    \frac{\rho_\mu}{2} \int d\mathbf{r} h_{\nu \mu}(r) [h_{\nu
      \mu}(r) - c_{\nu \mu}(r)] - \rho_\mu \hat{c}_{\nu
      \mu}(k=0)\right\}.
\end{align}
Note that these expressions hold only within the HNC approximation,
and are consistent with the virial route \eqref{eq:13}.

\section{Scaling properties}
\label{sec:scaling-properties}

As mentioned earlier, suitably reduced equilibrium properties of the
binary GCM depend on the five dimensionless combinations
$R_{12}/R_{11}$, $R_{22}/R_{11}$, $\epsilon^*_{11}$,
$\epsilon^*_{12}$, and $\epsilon^*_{22}$. The compressibility version
of the RPA thermodynamics (or equivalently, the mean field
approximation) allows a considerable reduction of this parameter
space. From the expressions \eqref{eq:10}-\eqref{eq:12c} it is clear
that the radii $R_{\mu \nu}$ and energies $\epsilon_{\mu \nu}$ enter
only in the combinations $V_{\mu \nu}$ defined in eqn. \eqref{eq:11}.
In terms of the reduced density $\rho^* = \rho V_{11}$ and pressure
$P^* = \beta P V_{11}$, the thermodynamic behaviour of the mixture is
uniquely determined by the dimensionless ratios $V_{22}/V_{11}$ and
$V_{12} / \sqrt{V_{11} V_{22}}$. This remarkable reduction of
parameter space holds only for the thermodynamics, but not for the
correlation functions, which depend explicitly on the five parameters
of the GCM potentials. The scale invariance of the thermodynamics is
broken by the RPA virial route, as is immediately evident form the
explicit expressions \eqref{eq:14} and \eqref{eq:15} (valid for a
symmetric mixture); the same is true within the HNC approximation.

However deviations from RPA-C (mean-field) scale invariance are
expected to be small in the high density, high temperature regime,
where RPA is increasingly accurate, so that the considerable reduction
in the number of relevant potential parameters (from 5 to 2) is
expected to carry over, at least approximately, to the phase diagrams
calculated with RPA-V or HNC thermodynamics, which will be presented
in the next section.

\section{Phase diagrams}
\label{sec:phase-diagrams}

The phase behaviour of the binary GCM may be deduced from the
knowledge of the reduced free energy per particle $f$ as a function of
the variables $v = 1/ \rho$ and $x$ and from its thermodynamic
derivatives (the pressure and the chemical potentials). For suitable
values of the potential parameters $\epsilon_{\mu \nu}$ and $R_{\mu
  \nu}$, the binary GCM becomes unstable against demixing at
sufficiently high densities. In terms of the free energy per particle,
$f = f(v,x)$, the standard thermodynamic stability conditions of a
binary mixture read \cite{ref1}:
\begin{subequations}
  \label{eq:17}
  \begin{align}
    \left(\frac{\partial^2 f}{\partial v^2}\right)_x > 0; \qquad
    \left(\frac{\partial^2 f}{\partial x^2}\right)_v &>0\\
    \left(\frac{\partial^2 f}{\partial v^2}\right)_x
    \left(\frac{\partial^2 f}{\partial x^2}\right)_v -
    \left(\frac{\partial^2 f}{\partial v \partial x}\right)^2 &>0.\label{eq:17b}
  \end{align}
\end{subequations}
The first condition ensures mechanical stability, while the second
inequality guarantees stability against demixing at constant volume;
the third inequality ensures stability of the mixture at constant
pressure. Note that the stability at constant pressure is the more
restrictive condition.

\begin{figure}[htbp]
  \begin{center}
    \includegraphics[angle=-90,width=14cm]{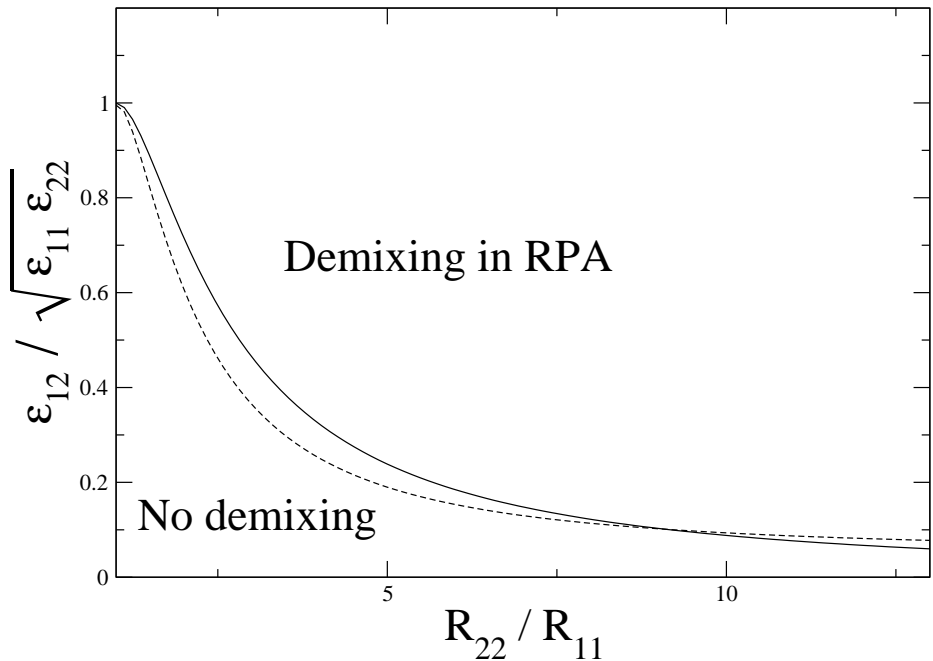}
    \caption{Mixing/demixing regions within RPA-C, in the
      $\epsilon_{12}/\sqrt{\epsilon_{11} \epsilon_{22}}$ vs.
      $R_{22}/R_{11}$ plane. Above the solid line phase separation is
      predicted by RPA-C. Corresponding values of
      $\epsilon_{12}/\sqrt{\epsilon_{11} \epsilon_{22}}$ predicted for
      polymers with radii of gyration $R_{11}$ and $R_{22}$ by a
      renormalisation group theory of Kr{\"u}ger {\em et al.} \cite{ref10} are
      plotted as the dashed curve. The intersection of these curves at
      $R_{22}/R_{11} \approx 10$ suggests the possibility of a phase
      separation in polymer mixtures with extreme size ratios.}
    \label{fig:2}
  \end{center}
\end{figure}

The subsequent discussion is
restricted to demixing at constant pressure. The vanishing of the
l.\ h.\ s.\ of equation \eqref{eq:17b}, corresponding to the case where the third inequality turns
into an equality, signals the occurrence of spinodal
instability. Within the mean-field (RPA-C) approximation
\eqref{eq:10}, the equation for the spinodal is easily calculated to
be \cite{ref17}
\begin{equation}
  \label{eq:18}
  1 + \rho V_1 (x) - \rho^2 x (1-x) \Delta V = 0,
\end{equation}
where:
\begin{subequations}
  \label{eq:19}
  \begin{align}
    \label{eq:19a}
    V_1(x) &= (1-x) V_{11} + x V_{22}\\
    \label{eq:19b}
    \Delta V &= V_{12}^2 - V_{11} V_{22}.
  \end{align}
\end{subequations}
It is easily inferred from \eqref{eq:18} that a spinodal instability
occurs whenever $V_{12} / \sqrt{V_{11} V_{22}} > 1$. This region is
visualized in Figure~\ref{fig:2} in a plot of the minimum energy ratio
$\epsilon_{12} / \sqrt{\epsilon_{11} \epsilon_{22}}$ required for
spinodal instability, as a function of the size ratio $R_{22} /
R_{11}$. The radius $R_{12}$ is taken to be given by the combination
rule:
\begin{equation}
  \label{eq:20}
  R_{12}^2 = \half (R_{11}^2 + R_{22}^2),
\end{equation}
as suggested by renormalization group (RG) calculations \cite{ref10}
and by direct simulation  \cite{ref11} of mixtures of self-avoiding polymer
coils.

Figure~\ref{fig:2} also shows the values of the ratio $\epsilon_{12} /
\sqrt{\epsilon_{11} \epsilon_{22}}$, as calculated for two
self-avoiding walk polymers by RG techniques
\cite{ref10}, plotted as a function of the ratio of their radii of
gyration. The figure suggests that phase separation of real polymers
would only be observed for ratios of gyration radii larger than 10,
i.\ e.\ for mixtures of very long and very short polymers. However, the
RG results are probably not trustworthy for such asymmetric mixtures,
and the RG potentials are strictly valid only in the infinite dilution
limit.

\begin{figure}[htbp]
  \begin{center}
    \includegraphics[width=14cm,angle=-90]{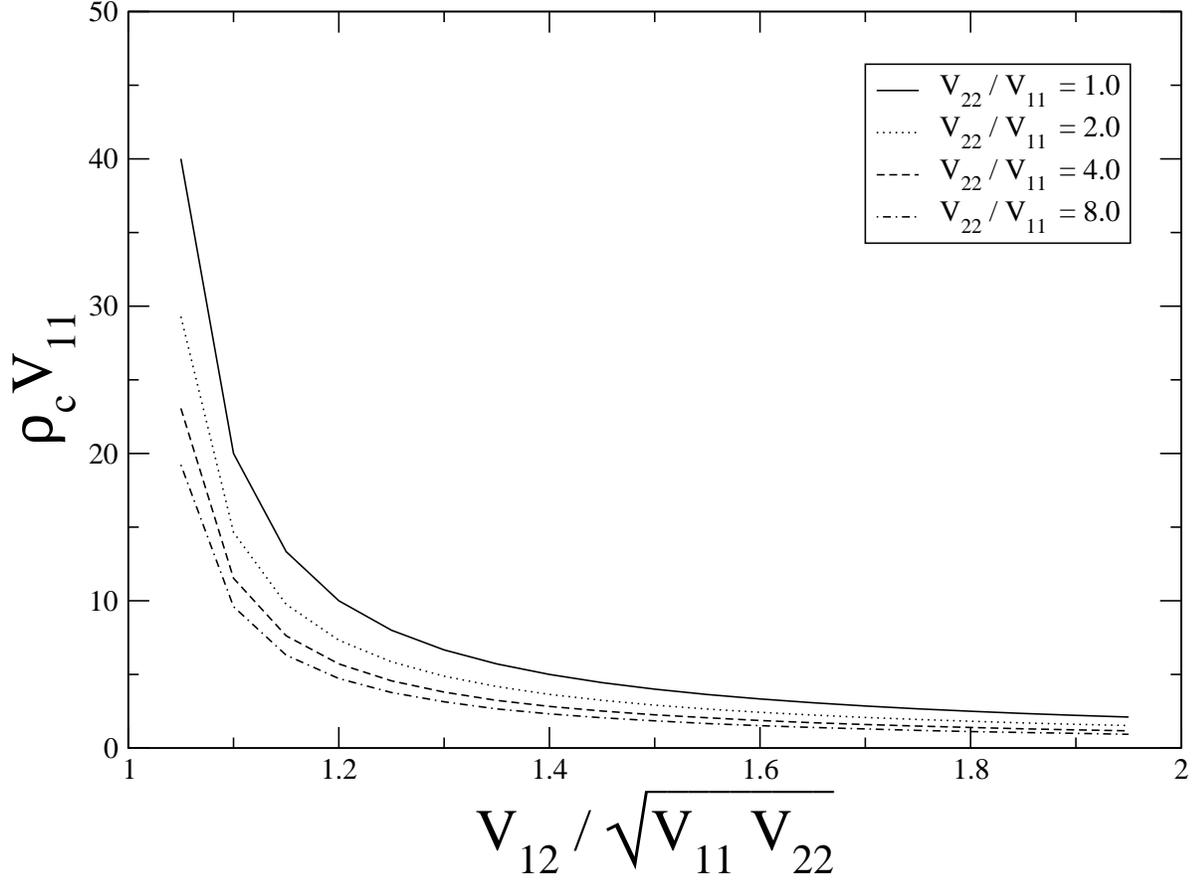}
    \caption{Reduced critical densities $\rho_c V_{11}$
      vs. $V_{12}/\sqrt{V_{11}V_{22}}$, for several values of
      $V_{22}/V_{11}$ from RPA-C (shown in the inset).}
    \label{fig:3}
  \end{center}
\end{figure}

The spinodal line in the $P-x$-plane is easily calculated from
\eqref{eq:18} to be:
\begin{equation}
  \label{eq:21}
  \rho_S(x) = \frac{V_1(x) + \sqrt{[V_1(x)]^2 + 4 x (1-x) \Delta V}}{2
  x (1-x) \Delta V}
\end{equation}
The critical concentration is determined by the condition:
\begin{equation}
  \label{eq:22}
  \frac{d P_S(x)}{dx} = 0,
\end{equation}
where $P_S(x)$ is the pressure calculated from \eqref{eq:12a}, at the
spinodal density given by \eqref{eq:21}. The corresponding critical
density $\rho_C$, calculated by substituting the critical
concentration determined by \eqref{eq:22} is plotted in
Figure~\ref{fig:3} versus the ratio $V_{12} / \sqrt{V_{11} V_{22}}$,
for several ratios $V_{22} / V_{11}$. 

\begin{figure}[htbp]
  \begin{center}
    \includegraphics[angle=-90,width=\textwidth]{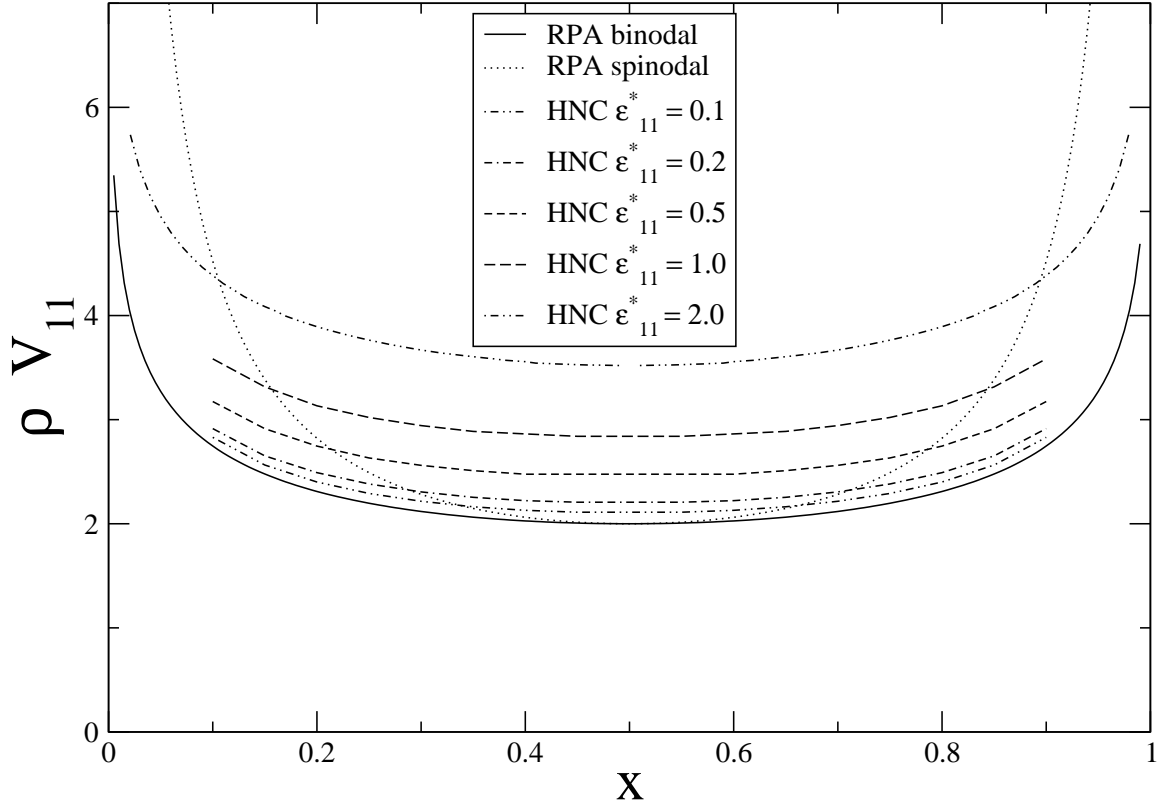}
    \caption{Binodal for a symmetric GCM mixture with $V_{12} /
      V_{22} = 2$ for values of $\epsilon^*_{11} = \epsilon^*_{22} = 0.2,
      0.5, 1.0, 2.0$, determined by RPA-C (full curves) and HNC
      (dashed curve). The RPA-C binodal and spinodal (dotted line)
      depend only on the ratio $V_{12}/V_{11}$. The HNC binodal does
      not obey the same scale invariance and is shifted to higher
      reduced density as $\epsilon_{11}$ increases.}
    \label{fig:4}
  \end{center}
\end{figure}

The binodal or phase coexistence curves in the $\rho-x$ plane are
determined by the usual conditions of equality of the chemical
potentials of both species and of the pressures in the two phases,
using expressions \eqref{eq:12a}-\eqref{eq:12c}.  The resulting
equations must in general be solved numerically to yield the binodal
curves.  The complete phase diagram can however be calculated
analytically within the RPA-C approximation, in the symmetric case,
where $V_{11} = V_{22}$. Due to the symmetry of the problem, all
thermodynamic quantities must be invariant with respect to the
transformation $1 \leftrightarrow 2; x \leftrightarrow (1-x)$ and
$\rho \leftrightarrow \rho$. This simplification allows the fully
analytic treatment detailed in Appendix B. The resulting spinodal and
binodal are shown in Figure~\ref{fig:4} in the case where $V_{12} /
V_{11} = 2$. The same Figure also shows the HNC binodals, for several
values of $\epsilon^*_{11}$ (remember that within RPA-C all these
binodals coincide).  As expected the HNC results break the scale
invariance of RPA-C and shift the binodal curves upward, to higher
densities. The HNC and RPA-C critical densities differ by almost a
factor of 2 for $\epsilon^*_{11} = 2$, pointing to the limitations of
the mean-field (RPA-C) description.

\begin{figure}[htbp]
  \begin{center}
    \includegraphics[angle=-90,width=\textwidth]{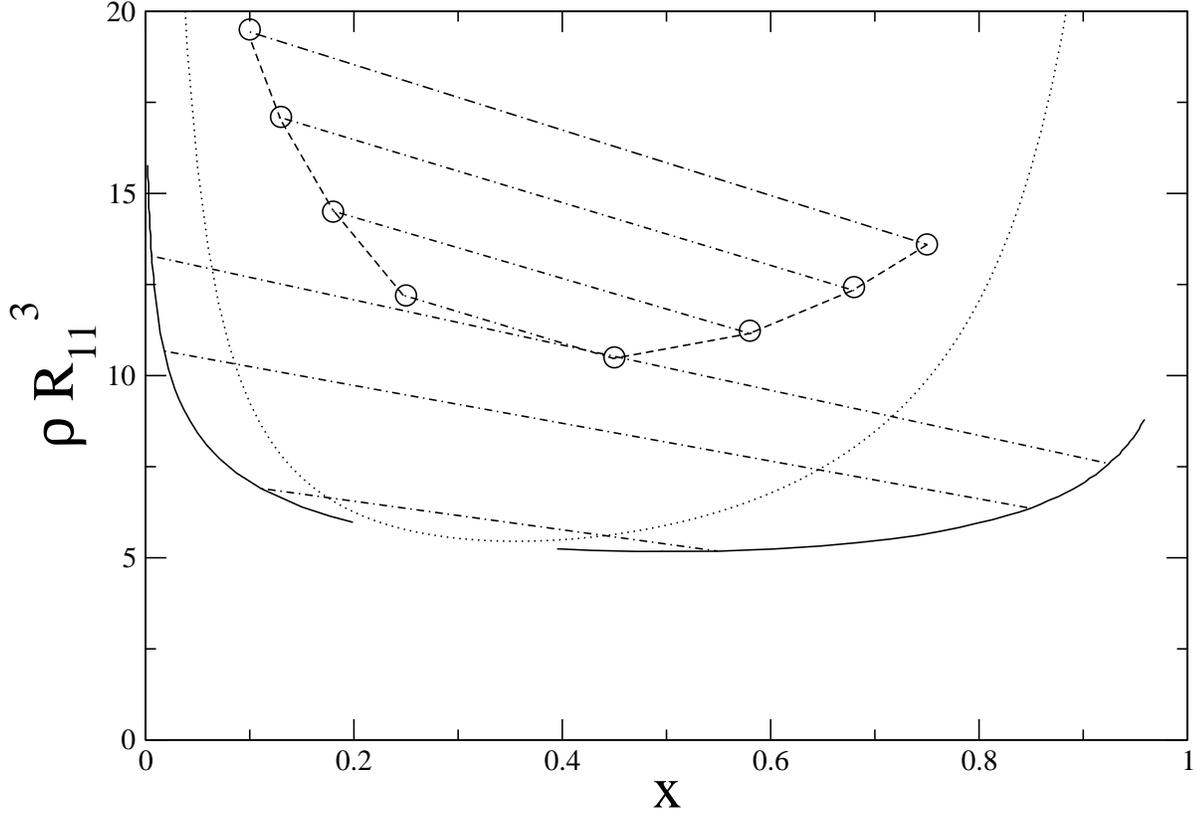}
    \caption{Phase diagram in $\rho-x$ plane for asymmetric mixture
      with $R_{11} = 1, R_{12} = 0.665, R_{22} = 0.849,
      \epsilon^*_{11} = \epsilon^*_{22} = 2, \epsilon^*_{12} =
      1.888$. The solid curve is the RPA-C binodal with examples of
      tie-lines of coexisting phases as dash-dotted lines. The dotted
      line is the RPA spinodal. The circles
      and dashed curve represent the HNC binodal with tie-lines shown
      again as dash-dotted lines.}
    \label{fig:5a}
  \end{center}
\end{figure}

\begin{figure}[htbp]
  \begin{center}
    \includegraphics[angle=-90,width=\textwidth]{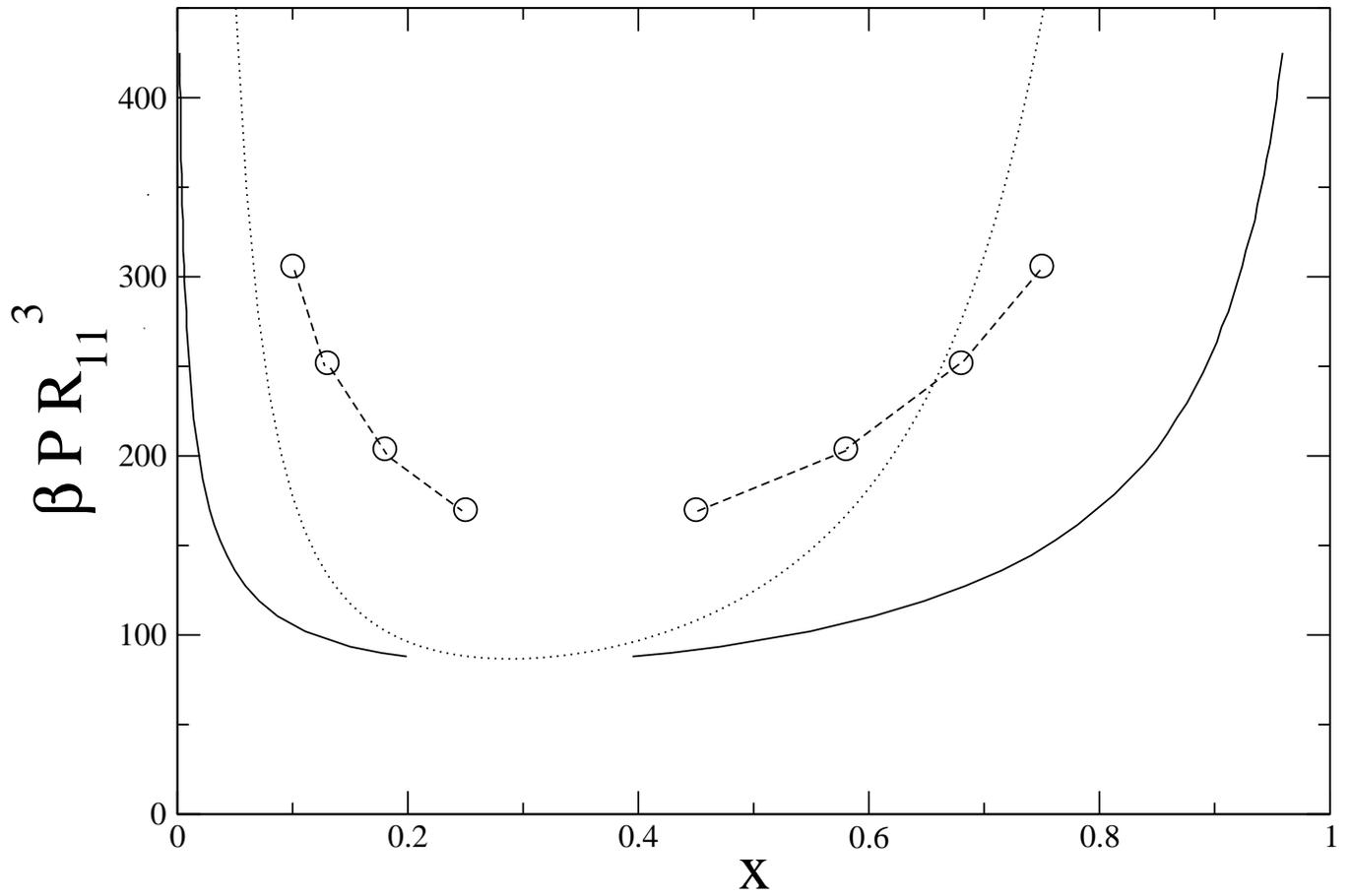}
    \caption{Phase diagram in $P-x$ plane for asymmetric mixture
      with $R_{11} = 1, R_{12} = 0.665$ and $R_{22} = 0.849,
      \epsilon^*_{11} = \epsilon^*_{22} = 2, \epsilon^*_{12} =
      1.888$. Symbols as in Fig. \ref{fig:5a}.  The tie-lines in this plot (not shown) are
      horizontal.}
    \label{fig:5b}
  \end{center}
\end{figure}

An example of a phase diagram in an asymmetric mixture, with potential
parameters already used in earlier RPA calculations
\cite{ref17,ref18}, is shown in Figure~\ref{fig:5a} in the $\rho-x$
plane and in Figure~\ref{fig:5b} in the $P-x$ plane. The spinodal and
binodal curves are now asymmetric, and the difference between RPA-C
and HNC results is again significant, with the HNC binodal being
pushed to higher densities.

\begin{figure}[htbp]
  \begin{center}
    \includegraphics[angle=-90,width=\textwidth]{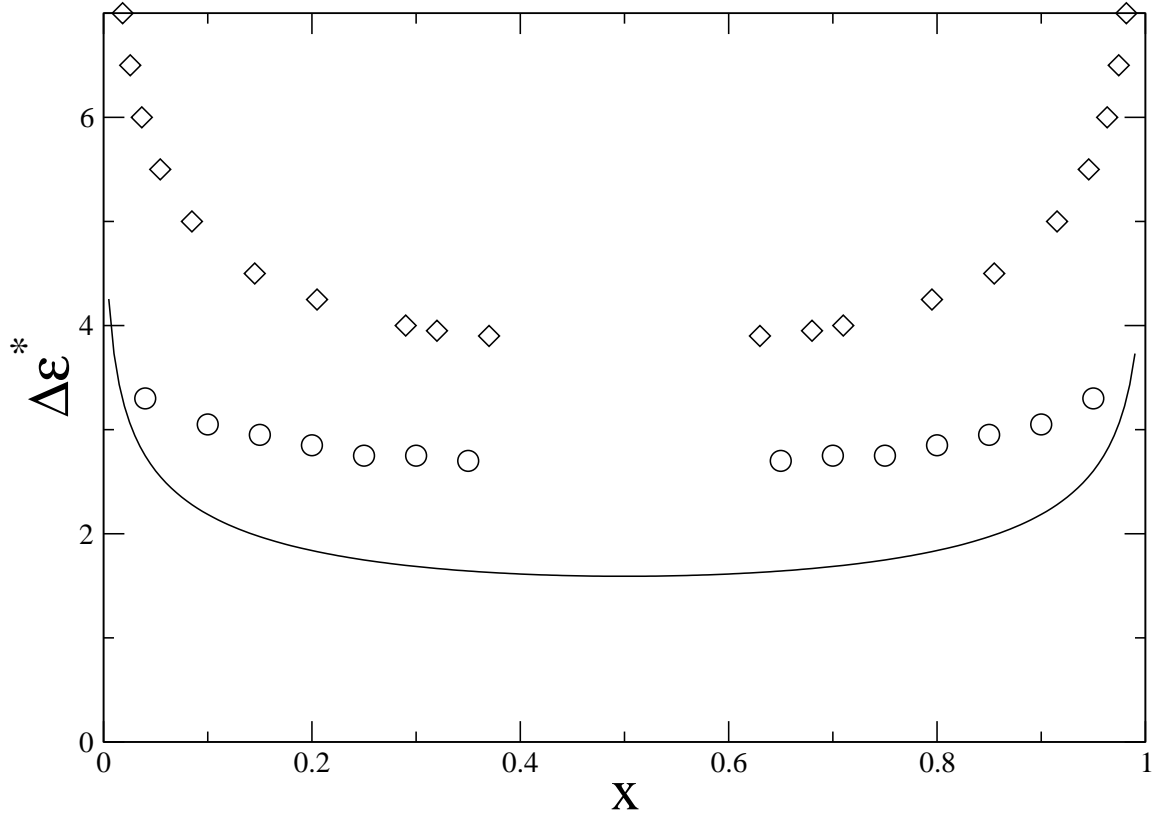}
    \caption{Binodal of the parabolic core system \eqref{eq:2} investigated by  Wijmans \textit{et
        al.} \cite{ref22}. Here $\epsilon^*_{11} = \epsilon^*_{22} =
      12.5$ is kept fixed and $\epsilon^*_{12}$ is determined so that
      the binodal density has the fixed value $\rho_b R^3 = 3.0$ at
      any given concentration $x$. The required difference $\Delta
      \epsilon^* = \epsilon^*_{12}-\epsilon^*_{11}$ is plotted vs. x.
      Note that in reference \cite{ref22} twice that value is plotted
      due to a different definition of the potential parameters. The
      diamonds correspond to the simulation data of Wijmans\textit {et
        al.} \cite{ref22}, the solid line represents the RPA-C
      prediction. The HNC results are plotted as circles. While RPA-C
      underestimates the required $\epsilon^*_{12}$ for the given
      binodal density by a factor of two, the HNC results lie closer
      to the simulation data, although a significant discrepancy
      remains. This may be due to the high amplitude $\epsilon^*_{12}$
      of the potentials, making the particles more hard-sphere like.}
    \label{fig:6a}
  \end{center}
\end{figure}

Phase separation in binary mixtures of particles interacting via the
closely related parabolic core potentials \eqref{eq:2} has been
observed in the Gibbs ensemble  Monte Carlo simulations of Wijmans et
al.\ \cite{ref22}. They considered the symmetric case where all
$R_{\mu \nu}$ are equal, $\epsilon^*_{11} = \epsilon^*_{22}$ and
$\epsilon^*_{12} = \epsilon^*_{11} + \Delta \epsilon^*$. We have
calculated the binodal in the $\Delta \epsilon^* - x$ plane under the
same conditions as the MC simulations, i.\ e.\ for $\epsilon^*_{11} =
12.5, \rho^*_b = \rho_b R^3 = 3$, in the RPA and HNC approximations. The
results are compared to the MC data in Figure \ref{fig:6a}. In view of
the fact that the amplitude of the parabolic repulsion is $12.5 k_B T$,
which makes it more hard-core like, the agreement may be considered to
be rather satisfactory. As in the GCM case, the HNC coexistence-curve
lies well above its RPA counterpart; it is closer to the simulation data. 

\section{Pair structure and clustering}
\label{sec:pair-struct-clust}

The pair correlation functions $h_{\mu \nu}(r)$ of the GCM mixture can
be calculated by combining the OZ relations \eqref{eq:6} with either
the RPA closure \eqref{eq:8} or the HNC closure \eqref{eq:9}. In the
symmetric case, characterized by a single gaussian range parameter
$R_{11} = R_{22} = R_{12} = R$, analytic expressions, in the form of
infinite series, can be obtained for the RPA closure, as shown in
Appendix A. In the asymmetric case, solutions for the $h_{\mu \nu}(r)$
are readily obtained by numerical Fourier transformation of equations
\eqref{eq:6} and \eqref{eq:7}, using the straightforward Fourier
transforms of the $c_{\mu \nu}(r)$ given by \eqref{eq:8}. 

The HNC closure requires an iterative solution of the coupled OZ and
closure equations. This was achieved using the standard Picard method,
and well-converged solutions were generally obtained in a few
iterations. Convergence was found to be slower in the vicinity of
phase coexistence, and to break down rapidly inside  the phase
coexistence region corresponding to metastable mixtures. Concentration
fluctuations build up in that region, as signalled by a $k = 0$ peak
of growing amplitude in the $\hat{h}_{\mu \nu}(k)$, or equivalently
in the corresponding partial structure factors
\begin{equation}
  \label{eq:23}
  S_{\mu \nu}(k) = x_\mu \delta_{\mu \nu} + x_\mu x_\nu \rho \hat{h}_{\mu \nu}(k)
\end{equation}
The concentration-concentration structure factor:
\begin{equation}
  \label{eq:24}
  S_{CC}(k) = x_2^2 S_{11}(k) - 2 x_1 x_2 S_{12}(k) + x_1^2 S_{22}(k)
\end{equation}
satisfies the long wavelength limit \cite{ref23}:
\begin{equation}
  \label{eq:25}
  \lim_{k \rightarrow 0} S_{CC}(k) = \frac{N k_B T}{(\partial^2 G /
  \partial x^2)_{N,P,T}},
\end{equation}
where $G$ is the Gibbs free energy.  The cross-over from metastable to
unstable mixture corresponds to the vanishing of $(\partial^2 G /
\partial x^2)_{N,P,T}$, so that $S_{CC}(k = 0)$ is expected to diverge
along a spinodal line. 

\begin{figure}[htbp]
  \begin{center}
    \includegraphics[angle=-90,width=\textwidth]{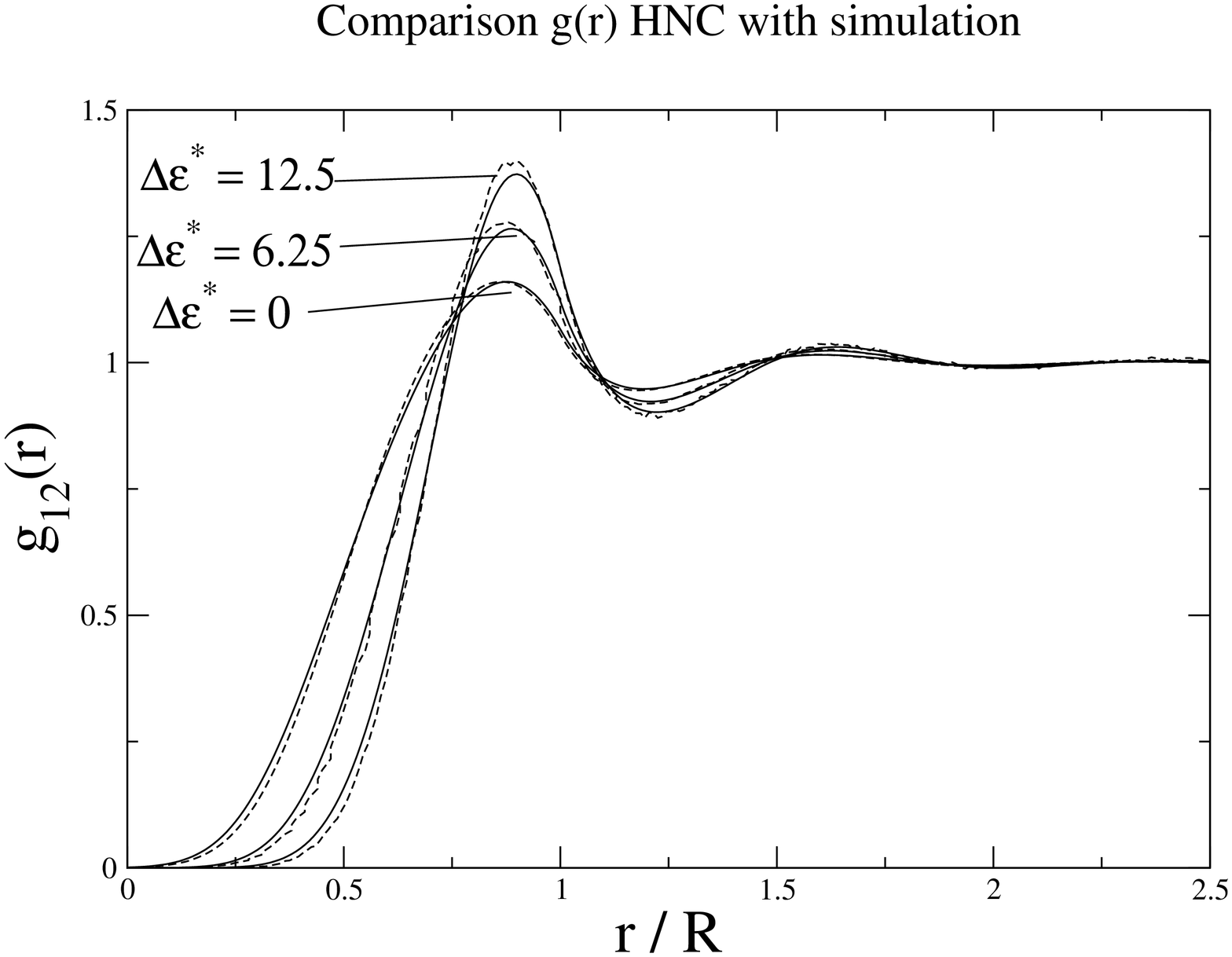}
    \caption{A comparison between HNC predictions for $g_{12}(r)$ and
      the simulation data of Wijmans \textit{et.\ al.} for the
      parabolic potential \eqref{eq:2} with $\epsilon^*_{11} = \epsilon^*_{22} =
      12.5$. The potential amplitude between unlike partikles
      $\epsilon^*_{12}$ is varied as  $\epsilon^*_{12} =
      \epsilon^*_{11} + \Delta \epsilon^*$ with $\Delta \epsilon^* = 0,
      6.25, 12.5$. The simulation data are plotted as 
      dashed lines, while HNC results are plotted as solid lines. Note
      that $g_{12}(r \rightarrow 0)$ is very near zero, in contrast to
      the case of the much softer potentials studied in \cite{ref17}
      and the subsequent Figures.}
    \label{fig:6}
  \end{center}
\end{figure}

This is indeed the case with the RPA closure, but the HNC closure
ceases to converge before a spinodal line is reached, a well-known
shortcoming of the HNC closure \cite{ref25}. Such a break-down is
typical of the inadequacy of fluid integral equations to describe
critical behaviour \cite{ref25a}. The accuracy of HNC in the stable
one-phase region for the pair correlation function $h(r)$, already
documented in the one-component GCM \cite{ref16,ref17}, is tested
under more severe conditions (namely $\epsilon^* = 12.5$) in
Figure~\ref{fig:6} against recent Monte Carlo data  \cite{ref22} for the parabolic
core potential \eqref{eq:2}.
The agreement is seen to be excellent. This suggests that the HNC
closure would be very useful to generate pair structure in
coarse-grained DPD fluids at a moderate computational cost.

\begin{figure}[htbp]
  \begin{center}
    \caption{Pair correlation function $h_{\mu \nu}(r)$ for the non-symmetric system $R_{11} =
      1.0, R_{12} = 1.582, R_{22} = 2.0,\, \epsilon^*_{11} =
      \epsilon^*_{12} = \epsilon^*_{22} = 2.0,\, \rho R_{11}^3 = 0.188, x =
      0.5$. The solid lines represent the RPA results, while HNC
      correlation functions are the dashed curves.}
    \includegraphics[angle=-90,width=12cm]{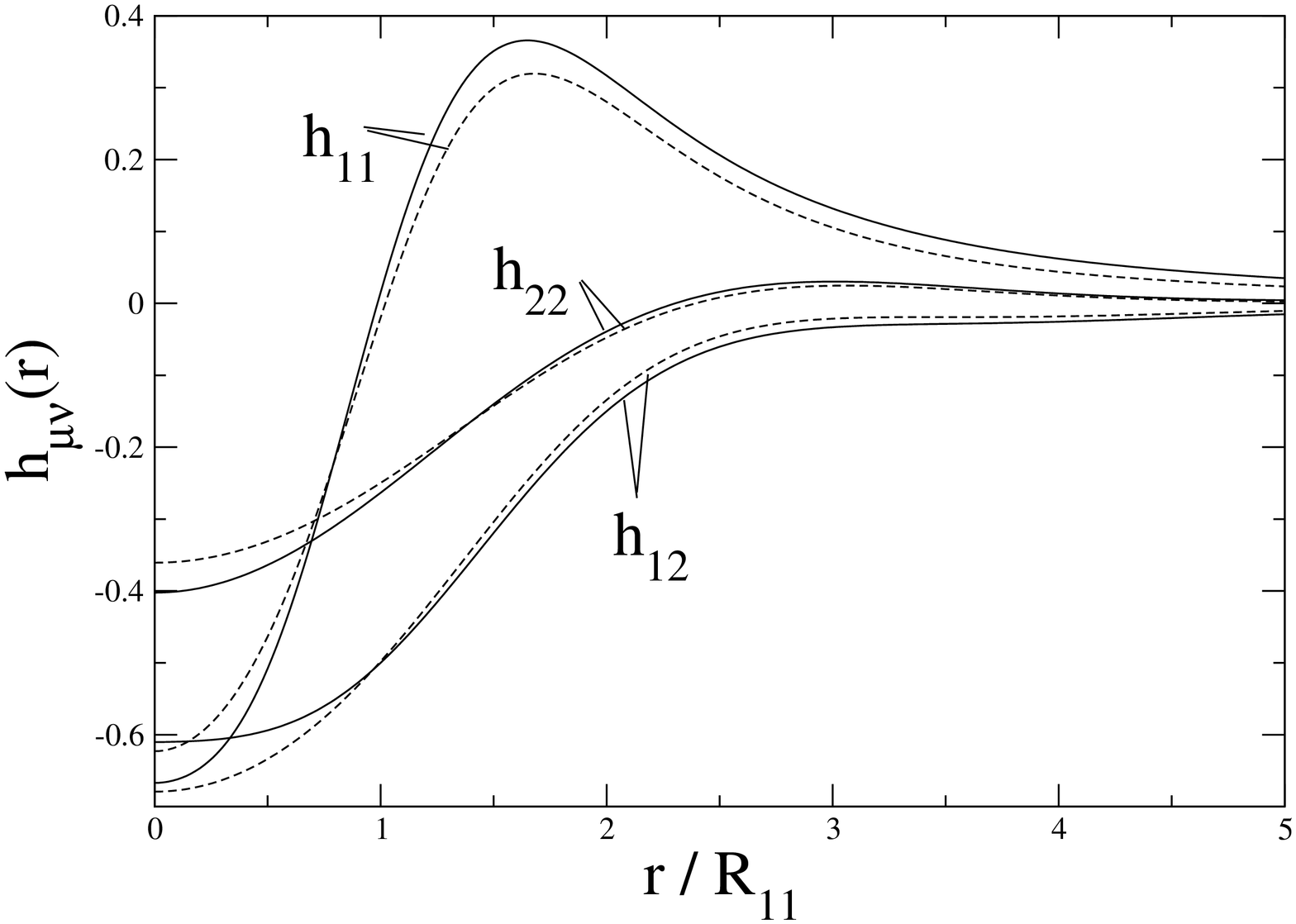}
    \label{fig:7}
  \end{center}
\end{figure}

\begin{figure}[htbp]
  \begin{center}
    \includegraphics[angle=-90,width=12cm]{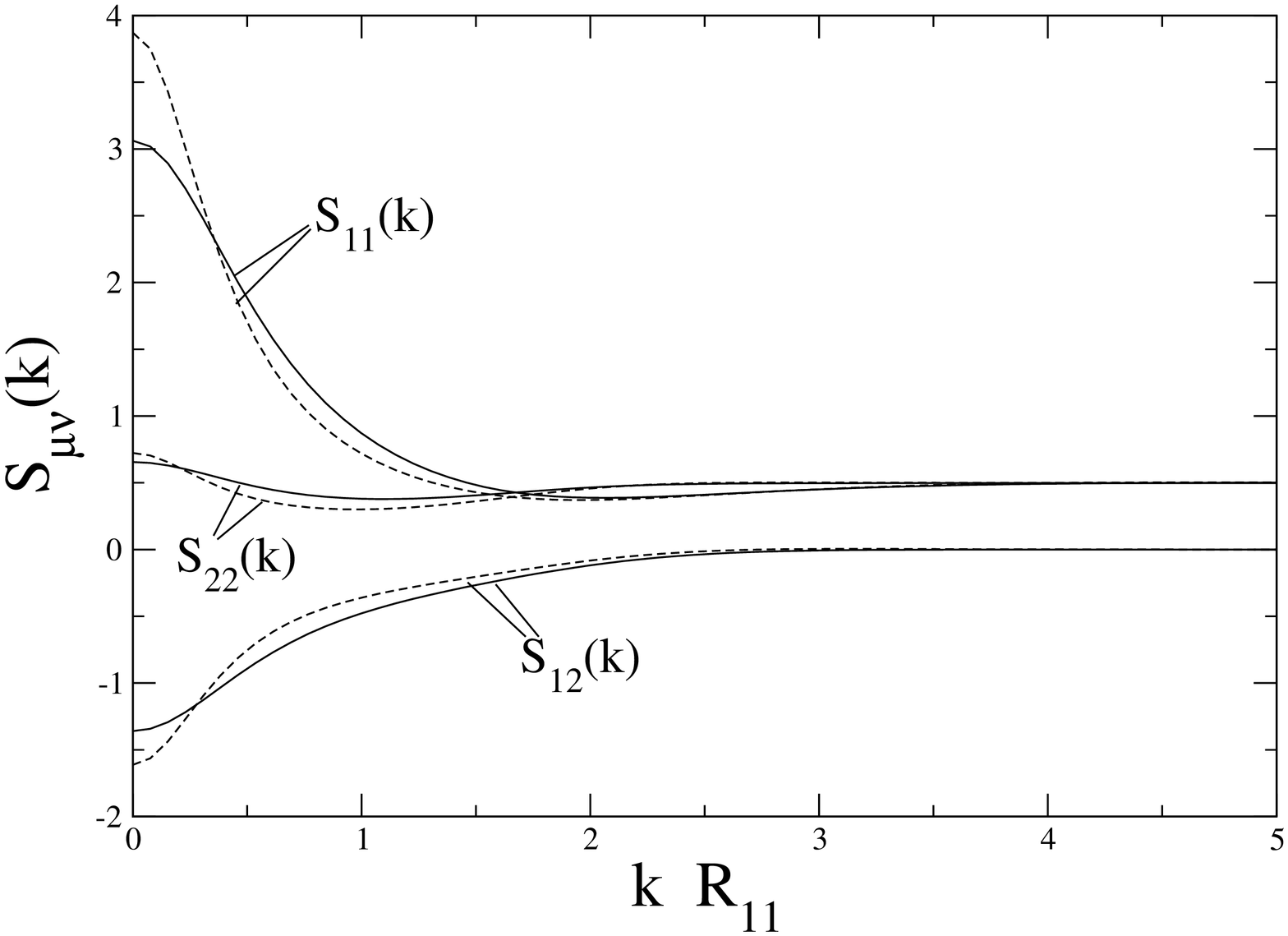}
        \caption{HNC partial structure factors for the non-symmetric system $R_{11} =
          1.0, R_{12} = 1.582, R_{22} = 2.0,\, \epsilon^*_{11} =
          \epsilon^*_{12} = \epsilon^*_{22} = 2.0,\, \rho R_{11}^3 =
          0.188, x = 0.5$. The solid lines represents the RPA partial
          structure factors, the dashed lines the HNC results. The
          enhanced values at $k=0$ show that this system is close to the
          spinodal.}
    \label{fig:8}
  \end{center}
\end{figure}

\begin{figure}[htbp]
  \begin{center}
    \includegraphics[angle=-90,width=\textwidth]{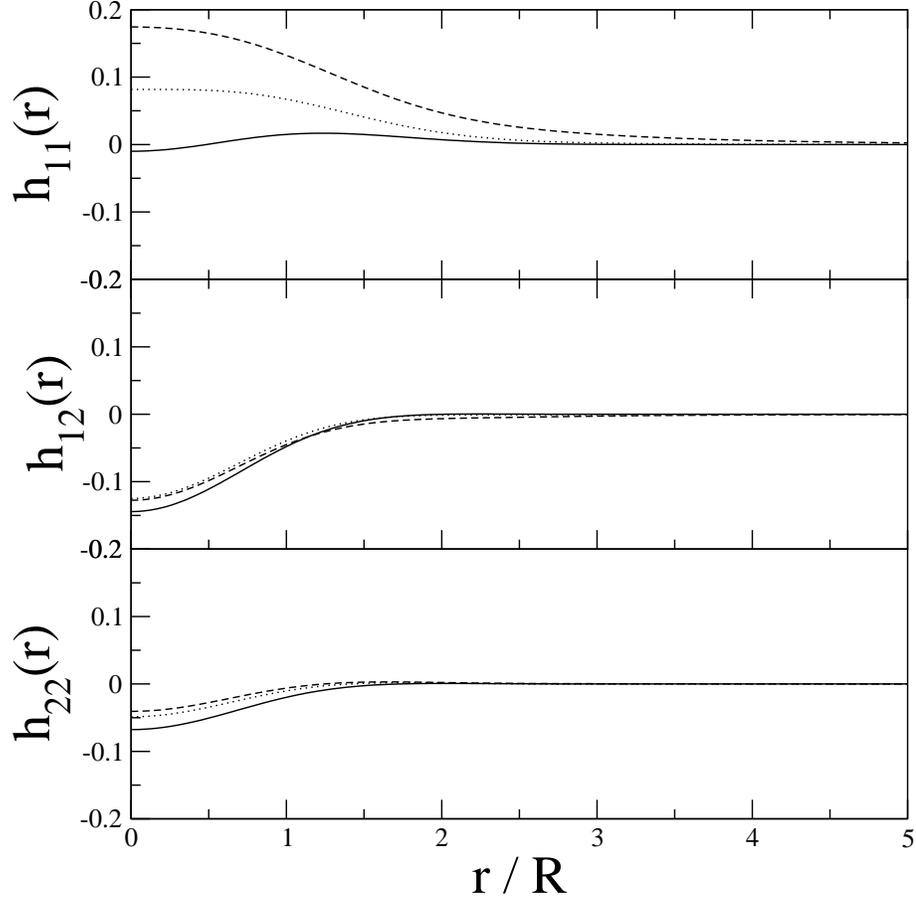}
    \caption{HNC pair correlation functions of particles in the symmetric
      case with $\epsilon^*_{11} = \epsilon^*_{22} = 0.1,
      \epsilon^*_{12} = 0.2$, at a concentration $x=0.9$ and densities below the
      binodal ($\rho R^3 = 2.0$, solid line), on the binodal ($\rho
      R^3 = 5.0$, dotted line) and near the spinodal ($\rho R^3 =
      6.7$, dashed line). The results for the minority
      species 1 point towards clustering.}
    \label{fig:9}
  \end{center}
\end{figure}

Examples of pair correlation functions and partial structure factors
for several states of GCM mixtures are shown in
Figures~\ref{fig:7}-\ref{fig:9}. The most striking feature is the
amount of structure observed both in $r$ and in $k$-space, even within
the RPA, compared to the previously studied one-component case
\cite{ref17}. The peak in the structure factors at $k=0$ as the
spinodal is approached is expected, as explained earlier. More
surprising perhaps is the appearance of peaks in the pair correlation
functions as $r \rightarrow 0$, or at finite $r$. Peaks at $r
\rightarrow 0$ are of course precluded in the presence of hard cores,
but are quite significant in the present penetrable core model and
point to a novel physical mechanism to trigger phase separation,
namely the clustering of the minority species, as observed e.\ g.\ in
Figure \ref{fig:9} for the symmetric case. The clustering can be
explained by a simple ``energetic'' argument: the particles of the
majority species tend to maximise their mutual distances. If the
repulsion between unlike particles is stronger than between like ones,
particles of the minority species prefer to cluster in the voids left
by the majority species, rather than overlap with particles of the
latter. This mechanism drives phase separation.

\section{Conclusion}
\label{sec:conclusion}
The two-component extension of the gaussian core model leads to
non-trivial phase behaviour, which we have investigated within RPA and
HNC theories. The former is partially analytic, while the latter requires
only a modest numerical effort to calculate partial pair correlation
functions and the resulting thermodynamic properties. While the pair
structure shows some unexpected features, the most interesting
prediction is the occurrence of phase separation induced by purely
repulsive pair interactions. As expected for this model of penetrable
particles, phase separation is driven by an enhanced repulsion between
unlike particles compared to the like-particle interaction.  

A special feature of the GCM mixtures is that at high densities
of both species, the mixture behaves like a ``mean field fluid'', i.\ 
e.\ the RPA becomes asymptotically exact \cite{ref16,ref17}. This
means that at finite densities, the RPA makes semi-quantitavely valid
predictions for the phase behaviour, and allows a rapid exploration of
potential parameter space to search for likely conditions for phase
separation.


The nature of the link between the behaviour of the binary GCM and
that of a binary polymer solution must still be worked out in detail,
mainly because the effective interactions between the centres of mass
of the polymer coils are state-dependent \cite{ref12}.  For a
one-component system at lower densities, the structure is not very
sensitive to the state-dependence \cite{Bolh01a,Bolh01b}.  We
therefore expect the binary GCM to make qualitatively correct
predictions for the structure of binary polymer solutions in the
dilute and beginning of the semi-dilute regimes.  The link to the
phase-behaviour is less clear.  Because the GCM uses a fixed
potential, it is mainly relevant to behaviour in the dilute regime.
While polymers in a melt are known to phase-separate rather easily
\cite{ref2a}, in the dilute regime for good solvents they are not
expected to phase-separate \cite{ref2b}.  This appears to be confirmed
by the results shown in Fig 2.  However, it may be that for poorer
solvent, or for stronger polymer incompatibilities, a regime where the
polymers phase-separate at low densities could open up.  If this is
the case, then the binary GCM would be a useful coarse-grained model
with which one could rapidly explore the qualitative phase behaviour
of this regime.

Another open question relates to the critical behaviour of the binary
GCM in the immediate vicinity of the critical consolute point. The RPA
clearly predicts mean field exponents.As regards the exact exponents,
it is not obvious whether the correct critical behaviour of the model
necessarily belongs to the Ising universality class, since there is no
clear correspondence between GCM mixtures and standard lattice models,
due to the penetrability of the particles. The question of the correct
universality class, a subject dear to the heart of Michael Fisher,
must be considered as open.

\section{Acknowledgements}
\label{sec:acknowledgements}

R. Finken is indebted to the Oppenheimer Fund for their
support. A. A. Louis would like to thank the Isaac Newton Trust, Cambridge, for
funding. 

\newpage
\begin{appendix}
\section{Solution of the virial route in RPA}
\label{sec:solut-viri-route}

We consider the GCM where the potentials have
equal range $R_{11} = R_{12} = R_{22} = R$, and the heights of the
repulsion potentials of like particles are the same for both species
$\epsilon_{11} = \epsilon_{22}$. Without loss of generality we can set
$V_{11} = 1$. Then we have
\begin{equation}
  \label{eq:b1}
  \hat{c}_{\mu \nu}(k) = - V_{\mu \nu} e^{-k^2 R^2 / 4},
\end{equation}
with
\begin{equation}
  \label{eq:b2}
  V_{\mu \nu} = \epsilon^*_{\mu \nu} R^3.
\end{equation}
If we introduce the abbreviation $\hat{z}(k) \equiv \rho V_{11} e^{-k^2
  R^2/4}$ and set $\Delta V = V_{12}^2 - V_{11}^2 = V_{12}^2 - 1$, the OZ equations \eqref{eq:6}
reduce to (using equation \eqref{eq:7}):
\begin{subequations}
\label{eq:b3}   
  \begin{align}
    \label{eq:b3a}
    \rho \hat{h}_{11}(k) &= \frac{ x_2 \Delta V \hat{z}(k)^2 - \hat{z}(k)}{\Delta(k)}\\
    \label{eq:b3b}
    \rho \hat{h}_{12}(k) &= - \frac{V_{12} \hat{z}(k)}{\Delta(k)}\\
    \label{eq:b3c}
    \rho \hat{h}_{22}(k) &= \frac{x_1 \Delta V \hat{z}(k)^2 - \hat{z}(k)}{\Delta(k)}\\
    \Delta(k) &= 1 + \hat{z}(k) - x (1-x) \Delta V \hat{z}(k)^2
    \label{eq:b3d}
  \end{align}
\end{subequations}
The denominator $\Delta(k)$ in equation~(\ref{eq:6}) has the zeros
\begin{subequations}
  \label{eq:b4}
  \begin{align}
    \label{eq:b4a}
    z_1 &= \frac{1 + \sqrt{1 + 4 x (1-x) \Delta 
        V}}{2 x (1-x)\Delta V} > 0\\
    \label{eq:b4b}
    z_2 &= \frac{1 - \sqrt{1 + 4 x (1-x) \Delta
        V}}{2 x (1-x)\Delta V} < 0
  \end{align}
\end{subequations}
The positive root $z_1$ is the spinodal density at any given
composition.

The virial equation \eqref{eq:13} gives
\begin{subequations}
  \label{eq:b5}
  \begin{align}
    \label{eq:b5a}
    \frac{\beta P^{\text{V}}}{\rho} - \frac{\beta P^{\text{C}}}{\rho}
    &= - \frac{2 \pi}{3} \rho \sum_\mu \sum_\nu x_\mu x_\nu \int_0^\infty r^3
    \frac{d \beta v_{\mu \nu}(r)}{dr} h_{\mu \nu}(r) dr\\
    &= \frac{1}{3} \rho \sum_\mu \sum_\nu x_\mu x_\nu
    \int d\mathbf{r} \epsilon_{\mu \nu} (r^2/R^2) e^{-r^2/R^2} h_{\mu \nu}(r).
  \end{align}
\end{subequations}
Evaluating this integral in Fourier space and making use of the Parseval theorem leads to
\begin{align}
  \label{eq:b6}
  \frac{\beta P^{\text{V}}}{\rho} - \frac{\beta P^{\text{C}}}{\rho}
  &= \frac{1}{3} \rho \sum_\mu \sum_\nu x_\mu x_\nu
  \int d\mathbf{k} V_{\mu \nu} \left[\frac{3}{2} - \frac{k^2
      R^2}{4}\right] e^{-r^2/R^2} \hat{h}_{\mu \nu}(k)\nonumber\\
  &= \frac{1}{12} \int d\mathbf{k} \frac{\hat{z}(k)^2}{\Delta(k)} [x_1
  x_2 \hat{z}(k) \Delta V - 2 x_1 x_2 \Delta V - 1] [6 - k^2 R^2] 
\end{align}
This integral cannot be evaluated analytically in general. However, if
the density is below the spinodal density for a given composition, one
can expand the denominator in the integrand in a power series with
respect to $\hat{z}(k)$. The integral can then be evaluated term by
term, interchanging the integration and the summation. In order to
make the symmetry between the two species explicit, we write $x =
\frac{1 + \xi}{2}$. The shifted concentration $\xi$ now varies between $-1$
and 1, with $\xi = 0$ corresponding to $x = 0.5$. 

With the definition of the polylogarithmic function
\begin{equation}
  \label{eq:a12}
  \Li_\alpha(x) = \sum_{n=1}^\infty \frac{x^n}{n^\alpha},
\end{equation}
and the auxiliary function
\begin{equation}
  \label{eq:a16}
  \aleph(\alpha) = \frac{1}{2 \alpha} [\Li_{3/2}(-\alpha) - \Li_{5/2}(-\alpha)]
\end{equation}
evaluation of \eqref{eq:b6} leads to
\begin{equation}
  \label{eq:a18}
  \begin{split}
    \frac{\beta P}{\rho} &= 1 +  \frac{1}{4}\rho [1 +  V_{12}   - 
    \xi^2 \Delta V]\\
    &-\frac{z_2 (1-\xi^2)\Delta V}{4 \pi^{3/2} R^3}
    \aleph\left(-\frac{\rho}{z_1}\right)\\
    &-\frac{z_1 (1-\xi^2)\Delta V}{4 \pi^{3/2} R^3}
    \aleph\left(-\frac{\rho}{z_2}\right).
  \end{split}
\end{equation}
This can be integrated to give the free energy per particle
\begin{equation}
  \label{eq:a19}
  \begin{split}
    \beta f(x,\rho) &= f^{\text{id}} + f^{\text{mix}}(x) \\
    &+ \frac{1}{4}\rho [1 +  V_{12}   - 
    \xi^2 \Delta V]\\
    &+ \frac{1}{2 \pi^{3/2} \rho R^3}[\Li_{5/2}(\rho/z_1) +
    \Li_{5/2}(\rho/z_2) + \rho]
  \end{split}
\end{equation}
From that we get the chemical potentials as
\begin{align}
  \beta \mu_1 &= \log(\Lambda^3 \rho)
  +\log\left(\frac{1-\xi}{2}\right) + \frac{1-\xi}{2} \rho + V_{12} \frac{1+\xi}{2}
  \rho + \frac{1}{2 \pi^{3/2} R^3}\nonumber \\
  &+ \frac{1}{2 \pi^{3/2} R^3 (1-\xi) \rho}
  \Li_{3/2}\left(\frac{\rho}{z_1}\right) \left[1 +
    \frac{ \xi}{\sqrt{1 + \Delta V (1-\xi^2)}}\right]
  \nonumber\\
  &+ \frac{1}{2 \pi^{3/2} R^3 (1-\xi) \rho} \Li_{3/2}\left(\frac{\rho}{z_2}\right) \left[1 -
    \frac{ \xi}{\sqrt{1 + \Delta V (1-\xi^2)}}\right],\\
  \beta \mu_2 &= \log(\Lambda^3 \rho) +
  \log\left(\frac{1+\xi}{2}\right) +   \frac{1+\xi}{2} \rho + V_{12} \frac{1- \xi}{2}
  \rho  + \frac{1}{2 \pi^{3/2} R^3}\nonumber\\
  &+ \frac{1}{2 \pi^{3/2} R^3 (1+\xi) \rho}
  \Li_{3/2}\left(\frac{\rho}{z_1}\right) \left[1 -
  \frac{ \xi}{\sqrt{1 + \Delta V (1-\xi^2)}}\right]
  \nonumber\\
  &+ \frac{1}{2 \pi^{3/2} R^3 (1+\xi) \rho}
  \Li_{3/2}\left(\frac{\rho}{z_2}\right) \left[1 +
    \frac{ \xi}{\sqrt{1 + \Delta V (1-\xi^2)}}\right].
\end{align}
It is easily checked that $\mu_1(\xi,\rho)=\mu_2(-\xi, \rho)$, due to
the symmetry of the model. Note that these expressions become singular
when $\rho \ge z_1 \equiv \rho_s$. This is a mathematical artefact
originating in the interchange of summation and integration in
\eqref{eq:b6}, which renders the expressions for $P$ and $\mu$ 
unusable for binodal calculations.

\newpage 
\section{Analytic solution of the RPA-C binodal in the symmetric case}
\label{sec:analyt-solut-rpa}
The phase diagram simplifies considerably in the symmetric case where
$V_{11} = V_{22}$.  We express $V_{12}$ as $V_{12} = V_{11} (1 +
\gamma)$ and note that demixing occurs when the dimensionless
parameter $\gamma$ is positive. As above we introduce the reduced
quantities ${\rho^*} = \rho V_{11}$ and ${P^*} = \beta P V_{11}$.
Because of the symmetry of the problem, all quantities must be
invariant with respect to the transformation $\rho \rightarrow \rho,\,
x \rightarrow 1 - x, 1 \leftrightarrow 2$, which corresponds to a
relabeling of species 1 and 2. As in appendix
\ref{sec:solut-viri-route}, we make use of the shifted concentration
$\xi$ to make this symmetry explicit. The pressure and the chemical potentials
can now be expressed in terms of the variables $\rho^*$ and $\xi$:
\begin{subequations}
  \label{eq:a1}
  \begin{align}
    \label{eq:a1a}
    P^* &= {\rho^*} + \frac{1}{4} {\rho^*}^2 (2 + \gamma) -
    \frac{1}{4} {\rho^*}^2 \gamma
    \xi^2\\
    \label{eq:a1b}
    \beta \mu_1 &= \ln({\rho^*}) + \ln\left(\frac{1-\xi}{2}\right) +
    \half {\rho^*}
    (2+\gamma) + \half {\rho^*} \xi \gamma\\
    \label{eq:a1c}
    \beta \mu_2 &= \ln({\rho^*}) + \ln\left(\frac{1+\xi}{2}\right) +
    \half {\rho^*} (2+\gamma) - \half {\rho^*} \xi \gamma
  \end{align}
\end{subequations}
In these equations we have set $\Lambda = V_{11}$, which merely shifts the free
energy per particle by a constant amount and thus does not change the phase
diagram.

Two coexisting phases $\alpha, \beta$ must have the same pressure and chemical
potentials, respectively:
\begin{equation}
  \label{eq:a2}
  P^{\alpha} = P^{\beta}, \; \mu_1^\alpha = \mu_1^\beta,\;
  \mu_2^\alpha = \mu_2^\beta
\end{equation}
Because of the symmetry in this model, coexisting phases must have the
same density $\rho^\alpha = \rho^\beta$ and the corresponding concentrations must fulfill
$x^\alpha = 1 - x^\beta$. Considering this, equal pressure in both phases is
trivially fulfilled, and the two conditions of equal chemical potentials become
equivalent. The relation between the binodal density $\rho_b$ and
concentration is 
\begin{equation}
  \label{eq:a3}
  {\rho^*}_b = \frac{1}{\gamma \xi} \ln\left(\frac{1+\xi}{1-\xi}\right).
\end{equation}
Expanding the binodal densities around the critical point gives
\begin{equation}
  \label{eq:a4}
  {\rho^*}_b =\frac{2}{\gamma}\left[1+\frac{1}{3}\xi^2 + \mathcal{O}(\xi^4)\right].
\end{equation}
From the series expansions we see that for all $\gamma >0$ the spinodal lies
inside the binodal. Within the RPA approximation we thus get an analytically
exact solution for a fluid-fluid demixing transition of a liquid consisting of
purely repulsive particles. As expected in this treatment, the resulting
critical exponents are mean-field exponents.
\end{appendix}

\newpage

\end{document}